# Energy conversion processes with perovskite-type materials

Davide Ferri, Daniele Pergolesi and Emiliana Fabbri

___________________________________

Correspondence

Dr. D. Ferri

E-mail: davide.ferri@psi.ch

Dr. D. Pergolesi

E-Mail: daniele.pergolesi@psi.ch

Dr. E. Fabbri

E-Mail; emiliana.fabbri@psi.ch

Paul Scherrer Institut, Forschungstrasse 111, CH-5232 Villigen, Switzerland

*Abstract:* Mixed oxides derived from the perovskite structure by combination of A- and B-site elements and by partial substitution of oxygen provide an immense playground of physico-chemical properties. Here, we account for own research conducted at the Paul Scherrer Institute on perovskite-type oxides and oxynitrides used in electrochemical, photo(electro)chemical and catalytic processes aiming at facing energy relevant issues.

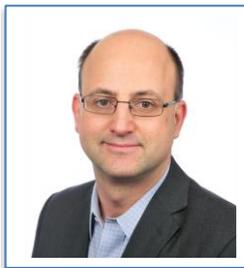

***Davide Ferri*** is currently group leader at the Paul Scherrer Institute and focuses research on catalysis and in situ/operando experiments using spectroscopy and diffraction methods. After studying industrial chemistry in Milano, he defended his doctoral degree at ETH Zurich in 2002 before making a short experience at Bruker Optics as sale manager and returning to ETH for a research assistant position. Before joining the Paul Scherrer Institute, he was group leader at Empa until 2012. Beside operando spectroscopy, he is interested in environmental catalysis, catalysis under unsteady-state conditions, the chemistry of perovskite-type oxides and catalysis in liquid phase.

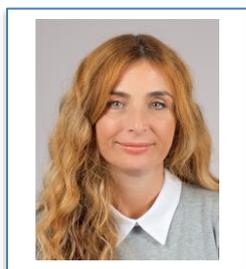

***Emiliana Fabbri*** is a research scientist at the Paul Scherrer Institute. Her main research interests are on materials development for electrochemical energy conversion devices. This includes materials fabrication along with electrochemical and operando X-ray spectroscopy investigations of surface states and electronic structures of catalyst materials towards a fundamental understanding of reaction mechanisms as well as design principles for new materials. Emiliana Fabbri completed her PhD in Materials Science in 2008 at the University of Rome Tor Vergata. Then, she held a scientist position at the International Center for Material Nanoarchitectonics (MANA) at the National Institute for Material Science (NIMS) in Japan. Since 2012, she has joined the Electrochemistry Laboratory at Paul Scherrer Institute, focusing on materials development for energy conversion devices.

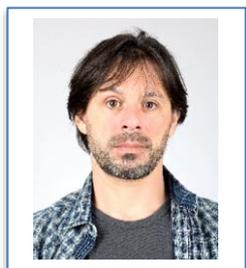

***Daniele Pergolesi*** is research scientist at the Paul Scherrer Institute. His research interests range from photo/electro-catalysis to solid-state ionic conduction using thin films as model systems to study fundamental physicochemical properties of materials. Daniele studied Physics and Material Science in Genova and was post-doctoral research fellow at the Italian National Institute for Nuclear Physics and at the University of Florida. Before joining the Paul Scherrer Institute, he was research scientist at the National Institute for Material Science in Japan from 2009 to 2013. Besides the field of electrochemical energy conversion and storage, Daniele is also interested in the fundamentals of the thin film growth process and strain engineering.

# 1. Introduction

Perovskite-type oxides are mixed metal oxides of formula $ABO_{3\pm\delta}$ where A is typically an alkaline, earth-alkaline or lanthanide element and B generally a transition metal. The perovskite crystal structure tolerates very large distortions with respect to the ideal $ABO_3$ cubic symmetry of $CaTiO_3$, the tilt angle of the $BO_6$ octahedra being able to account for large changes in the cation ionic radii. Moreover, large oxygen non-stoichiometry can be set to charge-compensate aliovalent substitutions. The result of these exercises is that a wide range of cations can be combined to form a wide range of perovskite-type compounds with all sort of material properties in optics, electronics, magnetism, multiferroicity, superconductivity, ionics, piezoelectricity, catalysis, etc.

The ease with which the oxygen non-stoichiometry can be stabilized allows fundamental modifications of the electronic structure with important impact for example on conduction properties, e.g. as in the well recognized electronic conductors $SrRuO_3$, Nb-doped $SrTiO_3$ and $SrTiO_{3-\delta}$. The creation of oxygen vacancies can allow oxygen ions to be the dominant charge carriers in the suitable gaseous environment and temperature range ($La_{1-x}Sr_xGa_{1-y}Mg_yO_{3-\delta}$).[1] In other perovskite-type compounds, the presence of oxygen vacancies makes it possible to uptake protons from the surrounding humidified gaseous environment through water dissociation, therefore opening proton conduction pathways (Y-doped $BaZrO_3$).[2] The large structural flexibility of perovskite-type materials allows also tailoring the defect chemistry in such a way that mixed ionic and electronic conductivities can coexist ($La_{1-x}Sr_xCo_{1-y}Fe_yO_{3-\delta}$).[3] While it is clear that electronic conductivity is essential for electrocatalysis, it has been recently proposed that also the ability of the perovskite structure to tolerate large oxygen deficiencies is essential for the electrochemical splitting of water.[4] Tuning electronic structure and conducting properties in perovskite-type materials is of fundamental importance for the development of next generation electrochemical energy conversion and storage systems, as well as for sensing and computing devices.

Catalytic activity of perovskite-type oxides is either intrinsic, for instance because of the presence of oxygen vacancies induced by various combinations of substitutions at both A- and B-sites or is induced by the presence of metal atoms and particles of a catalytically active element.[5] The active element can be incorporated at the A- or more typically at the B-site of the mixed oxide upon synthesis or is deposited in form of nanoparticles on the external surface of the oxide thus exploiting it as support with high degree of interaction. In the former situation, the perovskite-type oxide serves as precursor of the catalytically active species,[6] while in the latter these are directly available to the reactants.

In this work, we provide accounts of own research on the utilization of perovskite-type oxide materials in three different fields of applications at the Paul Scherrer Institut (PSI).

## 2. Perovskite oxides for the electrochemical water splitting

In the last decade, perovskites oxides have grabbed great attention as electrocatalysts for the oxygen evolution reaction (OER) in alkaline media. The OER is known to be the bottleneck reaction in the electrochemical splitting of water (i.e., water electrolysis) since it suffers from sluggish kinetics rendering high overpotentials at the anode side of water electrolyzers.

Water electrolyzers have emerged as a potential solution to mediate the intermittent generation of energy from renewable energy sources. Indeed, one of the main challenges related to the implementation of renewable energy technologies is related to their inconsistent energy flow, which can be achieved only by developing an efficient energy storage system. Water electrolyzers can offer an effective method to mediate the intermittent nature of renewable energies by using the energy surplus to produce hydrogen via water electrolysis. Different from the case of batteries, hydrogen as an energy vector can be stored outside the electrochemical device even for long term and used where and when needed. Therefore, large research efforts have been recently devoted to the development of efficient and durable water electrolyzers. The bottleneck reaction in water electrolyzers is the anodic reaction, such as the OER. State-of-the-art catalysts for the OER in alkaline environment are Ni-based oxides, which have been widely optimized in terms of OER activity over the past decades by Fe doping.[7] Recently, perovskite-type oxides have emerged as a novel class of promising electro-catalysts for the OER and studies have risen in the last years focusing on overcoming the current performance of Ni-based catalysts.[7a, 8]

Given the vast possible compositions allowed by the perovskite structure and the many open questions about the mechanism at the origin of their electrocatalytic activity towards the water splitting, many research efforts have been focused in the last decade on identifying activity descriptors able to predict the optimal perovskites physico-chemical properties for maximizing the OER activity. The Electrocatalysis and Interface group of Paul Scherrer Institut has actually achieved in the last five years a deep understanding of the OER mechanism on the surface of perovskite-type oxides proposing also design principles for a new generation of water splitting electrocatalysts. These remarkable results have been mainly achieved using *operando* techniques, such as *operando* X-ray absorption spectroscopy (XAS), small angle X-ray scattering (SAXS) and, near ambient pressure X-ray photoelectron spectroscopy (NAP-XPS)

In the literature most of the studies proposes a single activity descriptor, such that it is only a single physico-chemical property of the perovskite oxide family that determines their OER activity. The number of d-electrons, the $e_g$ band filling of the transition-metal cations, or the oxide formation energy have been proposed individually as activity descriptors for the electrocatalytic activity of perovskite-type oxides towards the OER.[9] In our recent work,[9] a wide range of perovskite compositions have been investigated experimentally and theoretically correlating several physico-chemical properties of the perovskites with their OER activity. In particular, the correlations between electronic conductivity, oxygen vacancy content and flat-band potential ($E_{fb}$) with the OER activity have been investigated. It has been found that, while all these parameters are crucial for the OER activity, none of them can be used alone for predicting the OER activity of perovskite-type oxides. An ideal correlation between one single physico-chemical property and the OER activity of a large pool of perovskite-type oxides never occurs but deviation points are always present.[10] This result suggests that such two-dimensional correlation (one physico-chemical property correlated to the OER activity) conceals strong limitations. However, our study [9] also shows that the deviation points present in a two-dimensional correlation can be explained considering other physico-chemical properties and their correlations with the OER activity. This systematic analysis led to a novel formulation of activity descriptors, where not only a single, but several physico-chemical properties of the perovskite catalysts are correlated with the OER activity resulting in a multi-descriptor correlation. The multi-descriptor correlation represents a significant advancement in the quest for highly active oxygen evolution catalysts for alkaline water electrolyzers.

A further development in the understanding of the perovskite surface dynamics under real operating conditions has been achieved by performing *operando* XAS for a series of nano-crystalline perovskite-type oxides, namely $Ba_{0.5}Sr_{0.5}CoO_{3-\delta}$ (BSC), $Ba_{0.5}Sr_{0.5}Co_{0.8}Fe_{0.2}O_{3-\delta}$ (BSCF), $La_{0.2}Sr_{0.8}CoO_{3-\delta}$ (LSC), $La_{0.2}Sr_{0.8}Co_{0.8}Fe_{0.2}O_{3-\delta}$ (LSCF) and $PrBaCo_2O_{5+\delta}$ (PBCO).[11] Using *operando* XAS measurements under OER conditions we have revealed, for the first time, that the perovskite electronic and local structure change during the water splitting reaction. Particularly, we have shown that the key for highly active perovskite OER catalysts is the formation of a self-assembled oxyhydroxide surface layer.[4a, 11a, 11c] This finding completely diverges from the more widely reported design principles for highly active perovskite-type catalysts,[7a] which generally proposed only ex situ perovskite properties as activity descriptors and have so far given no value to perovskite-type oxides developing this superficial amorphous layer.[4a]

In particular, by performing *operando* XAS on BSCF perovskite nano-structured oxide,[11a] which is a highly active and stable OER catalyst,[11b] it has been shown that the growth of a Co/FeO(OH) layer takes place only after the onset potential of the OER. The formation of highly active self-assembled Co/FeO(OH) on the surface of BSCF and the great performance stability of the BSCF-Co/FeO(OH) catalyst in alkaline environment have been explained by the reaction scheme depicted in Figure 1.[11a] This reaction scheme foresees the A-site cation dissolution driven by the lattice oxygen evolution reaction (LOER), an electrochemical process involving the direct evolution of lattice oxygen. In BSCF, Ba(II) and Sr(II) cations are highly soluble and, thus they can easily leach out from the perovskite structure according to the chemical equilibrium of Figure 1. The LOER can also trigger the dissolution of Co and Fe cations but, being the latter rather insoluble, Co and Fe re-deposition on the BSCF surface can take place leading to the formation of an amorphous Co/FeO(OH) surface layer. Furthermore, the lattice oxygen consumed by the LOER can be replenished by $OH^-$ present in the alkaline electrolyte solution. Therefore, a stable dynamic cycle is established, permitting the coexistence of a self-assembled active surface layer with the original BSCF perovskite structure.[4a, 11a]

The possibility that another oxygen evolution reaction mechanism can take place besides the conventional one, which foresees four concerted proton-electron transfer steps,[4a] has been proposed in the very early OER literature, but only recently it was rigorously formulated in our study.[12] Based on thermodynamics considerations, it is possible to demonstrate that if the conventional OER mechanism takes place, also LOER and metal cation dissolution are driven. A possible mechanism involving the oxidation of lattice oxygen (more precisely the lattice $O^{2-}$ anions) and named lattice oxygen evolution reaction (LOER) has been recently proposed by Fabbri and Schmidt. [4a]. The LOER, together with the conventional four concerted proton-electron transfer steps OER mechanism, is represented in Figure 2.

By performing *operando* XAS measurements on different Co-based catalysts, it was suggested that the more important the growth of the self-assembled oxyhydroxide layer is, the higher the OER/LOER activity of the catalyst. This also suggests that flexible perovskite structures such as BSCF are ideal precursor catalysts for the formation of a superficial self-assembled oxyhydroxide layer because their highly defective surface can facilitate oxygen exchange and dynamic self-reconstruction of the surface. Flexibility in the structure can be associated with the oxygen vacancy concentration and ion mobility, and indeed, a direct correlation between the oxygen vacancy content and the OER activity has been revealed in ref. [9].

In a recent study[13] we have deliberately produced Co/Fe oxyhydroxide layers on the surface of different supports, such as amorphous carbon and perovskite-type oxides. The study shows that an interplay between the oxyhydroxide and the support occurs, and particularly only perovskite supports with large oxygen vacancy content, e.g. BSCF, boost the Co-Fe oxyhydroxide OER activity. The model samples produced by intentionally depositing an oxyhydroxide phase on different supports suggests that perovskite-type oxides, which are ultimately able to build-up an active oxyhydroxide layer, are not barely a precursor of the active phase (the oxyhydroxide) but play also an essential role in boosting and sustaining the OER activity.[13] The same study also highlights the importance of the Fe presence in the self-assembled oxyhydroxide layer, being Co- or FeO(OH) catalysts much less active than the Co/FeO(OH) one.[13] Therefore, we have also carried out a systematic study to understand the functional role of Fe in perovskite OER catalysts with respect to activity and stability under operating conditions.[11c] In perovskite-type oxides as BSCF or LSCF, despite the fact that Fe constitutes only ~5 wt% of the mixed oxide, it greatly boosts both OER activity and stability performance. Particularly, the presence of Fe in BSCF and LSCF stabilizes Co in a lower oxidation state in the as prepared materials, maximizing the oxygen vacancy content, thus enabling the growth of the self-assembled oxyhydroxide layer on top of the perovskite structure under operating conditions.[11c] Furthermore, functional theory computational studies reveal that the presence of Fe enlarges the potential range of the thermodynamic metastability given by the Pourbaix diagrams, which in turns improves the stability of the material under operating conditions.[11c]

In summary, perovskite-type oxides have shown the capability of being active and stable electrocatalysts for water splitting in alkaline environment. Fundamental studies based on *operando* XAS measurements have revealed a novel reaction mechanism taking place on the oxide surface, the LOER, which leads to the formation of a highly active oxyhydroxide phase, mostly composed by the B-cations of the perovskite structure. However, the perovskite-type oxide is not simply a precursor for this oxyhydroxide active phase because only the co-existence of the perovskite structure underneath the oxyhydroxide layer provides high OER activity. Perovskite-type oxides in form of nano-powders have also shown remarkable performance in electrolyzers, exceeding the performance of the best OER catalyst used so far in the anodic electrode, e.g., $IrO_2$.[11a]

**3. Perovskite oxynitrides for sustainable solar fuel**

The use of solar energy to produce a clean and sustainable solar fuel is one of the most appealing strategy to meet our ever-increasing energy demand. A promising way to tackle this challenge relies on the photocatalytic water splitting, an artificial photosynthesis process where solar energy is used to decompose water molecules and generate $O_2$ and $H_2$ gas. As mentioned for the case of the electrochemical splitting of water, hydrogen can then be used as a renewable and sustainable energy vector, or solar fuel, available when and where is needed.[14] Achieving this goal is however a formidable task.

The solar water splitting process ($2H_2O \rightarrow 2H_2 + O_2$) relies on the availability of a semiconductor material with a band gap in the visible light energy range, therefore capable of absorbing photons from the sun, and photo-generate electron-hole pairs. These charge carriers should then possess enough mobility to reach the surface of the semiconductor in contact with water. If electrons and holes have energy higher than the hydrogen and oxygen redox potential respectively, and if they find enough catalytic activity at the solid-liquid interface, then the two particles can be used to promote the water splitting process.

Some perovskite-type materials are among the most promising visible light responsive semiconductors to tackle the challenge.[15] The electronic (and optical) properties of some metal oxides can be modified not only by engineering cationic substitution, but also changing the anionic chemical composition by substituting O with N in the lattice.[16] Many of the resulting oxynitride materials crystallize in the perovskite structure $ABO_{3-x}N_x$ (0<x<2). The A-site cation can be Ba, La, Ca, Sr or Y. The B-site transition metal cation is typically Ti, Ta, or Nb. Examples include $LaTiO_2N$, $SrTaO_2N$, $BaTaO_2N$, $CaNbO_2N$, $LaTaON_2$.[15, 17] These are semiconductor materials with the peculiar characteristic of a band gap in the range of ca. 1.8-2.5 eV, that is the visible light energy range. In addition, also the energy position of the band edges is particularly interesting for this class of materials. Setting at 0 V the hydrogen redox potential, the band gap of these materials include both the hydrogen and oxygen redox potentials. Therefore, these perovskite oxynitrides not only can absorb visible light photons (from the sun) and generate electron/hole pairs, but the photo-generated charge carriers possess the appropriate energy to promote the water reduction/oxidation reactions. Such properties make these perovskite oxynitrides very promising semiconductor materials for photoelectrochemical solar water splitting.[16a]

Powders of these materials are fabricated by thermal treatment of suitable precursor oxides in ammonia (ammonolysis) at temperatures ≥ 1000 ºC. Examples are the pyrochlores $Ca_2Nb_2O_7$ and $La_2Ti_2O_7$ that are used as the precursor oxides for $CaNbO_2N$ and $LaTiO_2N$, or the perovskites $LaTaO_4$ and $Ba_5Ta_4O_{15}$ that are used for $LaTaON_2$ and $BaTaO_2N$.[18] Figure 3(a) shows, as an example, the SEM micrograph of $LaTiO_2N$, where the ammonolysis process produces porous particles with grain size in the range of 600 nm up to 1.5-1.8 μm.[18a] It is important to note that the morphological features change significantly for different materials and the powder morphology has a critical impact on material functionality.

The precursor oxides typically have band gaps ≥ 3.5 eV, in the UV energy range. $TiO_2$ was the first material investigated for solar water splitting, and many other oxides have been characterized in the last decades. However, the relatively wide band gap of these oxide materials represents an intrinsic limitation for solar light-driven water splitting, since only a few percent of the solar spectrum is in the UV energy range. Therefore, shifting the photoresponse of the semiconductor down in energy is one of the main goals that must be achieved to make the photoelectrochemical process efficient. It is here that the perovskite oxynitrides listed above come into play. In these materials, the reduction of the band gap in the visible light energy range is a direct consequence of the N to O substitution into the pristine oxide. The oxygen 2p orbitals set the valence band maximum of the precursor oxides. The hybridization of the O and N 2p orbitals creates additional energy levels above the valence band maximum, therefore reducing the band gap by shifting upward in energy the valence band maximum. Recently, it was observed that for LaTiO$_2$N the N substitution not only shifts the energy of the valence band maximum upwards, but also moves the energy of the conduction band minimum downwards. Both these effects have to be taken into consideration to explain the overall reduction of the band gap.[19]

The specific chemical composition also affects the width of the band gap. Theoretical calculations of the band structures of ATaO$_2$N with A = Ba, Sr, and Ca predict larger band gap with increasing the orthorhombic distortion (SrTaO$_2$N and even more in CaTaO$_2$N) with respect to the cubic structure of BaTaO$_2$N.[16b] The B – O – B bond angle, which is 180º in the cubic structure, becomes smaller by decreasing the size of the A-site cation. Such a structural distortion leads to smaller width of the conduction band leaving the band energetic centre unaffected. The overall effect is a progressively larger band gap with increasing the lattice distortion. Simulation of the band structure of BaTaO$_2$N and BaNbO$_2$N[16b] concluded that the smaller electronegativity of the B-site cation (Ta compared to Nb) increases the width of the band gap by shifting the energetic centre of the conduction band upwards.

For photoelectrochemical (PEC) measurements, the oxynitride powders are "decorated" with appropriate nanoparticles of a cocatalyst, e.g. be IrO$_y$, Ni(Fe)O$_y$ or cobalt phosphate (CoPi).[18b] If the photo-generation of charge carriers and their migration to the surface of the semiconductors are both bulk properties, once arrived at the solid-liquid interface electrons and holes have to be extracted and used for the electrochemical reaction. This is the role of the cocatalyst nanoparticles; to scavenge the charge carriers and to passivate the surface.[20]

The conventional three-electrode configuration (anode, cathode, and reference electrode) is often used for PEC characterization towards light-driven water splitting. For this measurement, the anode is fabricated by electrophoretic deposition of the decorated powders on a conducting support (typically fluorine-doped tin oxide coated glass - FTO) that works as the current collector. Alternatively, the layer of powder of the semiconductor can also be first deposited on the substrate and then decorated with the cocatalyst. Finally, a thin coating of Ti or Ta oxide, fabricated by drop casting and annealing, is used to provide good electrical contact between particles and between particles and current collector. Figure 3(b) illustrates schematically the complete photoelectrode.

The sample is immersed into a suitable electrolyte and illuminated. The photogenerated holes reach the surface and are consumed for the oxygen evolution reaction, while the photogenerated electrons are transferred from the current collector to a Pt wire, the cathode electrode, placed in the same electrochemical cell, where $H_2$ evolves. The electronic current between anode and cathode, which is proportional to the amount of generated hydrogen, is then measured at different applied potentials with respect to a reference electrode that set the 0 V at the $H^+$/H redox potential. A schematic illustration of the PEC setup can be seen in Figure 3(c). To distinguish voltage- and light-driven effects, chopped illumination is applied to measure the "dark" and "light" current vs. applied bias.

PEC measurements based on this experimental setup allow not only the characterization of powder samples, but also that of thin films of perovskite oxynitrides, providing a conducting substrate is used as the growth platform.[21] Thin films can be grown with higher crystallinity and very good grain-to-grain and grain-to-current collector electrical contact. Therefore, the electron/hole recombination can be strongly suppressed compared to powder samples. In principle, considering comparable band gaps, i.e. assuming the same N content for film and powders, the bulk morphology of a thin film-based photoanode is expected to facilitate the PEC process.[18a] However, a powder-based photoanode is characterized by a surface roughness that for these materials can easily be 50 to 100 times larger than that of a thin film-based device. Another very important advantage of powder compared to thin film is related to the fabrication costs, which in the latter case are orders of magnitude larger.

In summary, a device for solar hydrogen production will certainly be fabricated with powders of the selected semiconductor, whether oxynitrides or other materials will depend on materials development. However, thin film based photoelectrodes are invaluable tools for fundamental investigations.

As mentioned above, the bulk properties of thin film semiconductors can be probed independently, to some extent, from the specific morphological features of the powders and the result of the necking treatment.[18a] In addition, and more importantly, the distinctive feature of a thin film is the very well-defined surface, often atomically flat. Thin films are indeed an ideal experimental platform for surface-sensitive studies of the solid/liquid interface, which is where the electrochemical reaction takes place.[21b]

Figure 4 shows a cross sectional TEM micrograph of a 150 nm thick $LaTiO_xN_y$ thin film grown on a (100)-oriented MgO single crystal substrate. A TiN buffer layer is used in-between substrate and film. Due to the very good crystallographic matching with the substrate, the buffer layer grows epitaxially oriented with good crystalline quality and provides an ideal seed layer for $LaTiO_xN_y$. TiN not only is a very good electronic conductor that can serve as the current collector for PEC characterizations, but it also provides a sort of N reservoir for the growing film. At the deposition temperature of about 750 °C, N and O can be exchanged between TiN and LTON. This does not affect the conducting properties of the buffer layer but it does affect the N content on the film. Comparing the N content for films grown on oxide materials and on TiN-coated substrates, in the latter case the N content can be higher and more reproducibly attainable.[21a]

LaTiO$_x$N$_y$ polycrystalline and epitaxial thin films, grown by pulsed reactive crossed-beam laser ablation with different orientations, were used to probe the charge transport and transfer properties in samples that interface with the electrolyte with different crystal planes and/or different chemical terminations. Figure 4 (a), (b), and (c) show the LaO, La-N, and TiO$_2$ terminations of the (001) surface, respectively for the energetically most preferential anion order.[21b] The absorbed photon-to-current conversion efficiency was used as the figure of merit to compare PEC performance. For the perovskite structure of LaTiO$_x$N$_y$, the (001)-oriented surface emerged as that with the higher electrochemical activity. Applying density functional theory, the surface structure of the perovskite oxynitride was calculated. La-N turned out to be the thermodynamically most stable surface termination, which is also the chemical termination that allows the easiest subsurface-to-surface charge transfer.

The natural follow-up of this investigation includes the study of different perovskite oxynitrides and the characterization of cocatalyst-decorated surfaces. The final target of this ambitious research path is the identification of a material (or combination of materials) with the best bulk optical and transport properties and the highest electrochemical activity for water splitting at the surface.

Severe materials issues have still to be addressed before H$_2$ can be generated as a clean solar fuel from water. Succeeding would represent the finding of the Holy Grail capable of meeting our energy demand in a sustainable way. The question would eventually be whether humanity will be able to handle a virtually unlimited energy source, the sunlight.

From the technological point of view, two basic reactor designs for photoelectrochemical hydrogen generation have been proposed.[22] One is based on a colloidal suspension in aqueous solution of decorated particles of the most appropriate visible light responsive semiconductor. The other is based on semiconducting particle coatings on rigid supporting structures to form panel arrays.

Several techno-economic analyses, comparing different hydrogen generation technologies in terms of efficiency, capital/operating costs, average lifetime, semiconductor manufacturing, etc., foresee encouraging potentials for the photocatalytic and/or photoelectrochemical solar water splitting.[22-23] However, the cost comparison with H$_2$ generated by low cost hydrocarbon reforming is inexorable and tell us clearly that only a consistent CO$_2$ tax added to the consumption of hydrocarbon resources would allow any kind of solar fuel to become competitive in the energy market.[23] The question is whether the cost comparison between solar and fossil fuel in an unconstrained $CO_2$ energy market can really be considered as a fair play.

**4. Ni-based catalysts for $CO_2$ hydrogenation to $CH_4$**

At PSI much research is devoted to the production of synthetic natural gas (SNG) from $CO/CO_2$ and $H_2$ feedstocks present after gasification of dry[24] and wet biomass.[25] The reason is that methane ($CH_4$), the major component of natural gas (NG) and SNG, displays the lowest C:H ratio thus producing the least $CO_2$ among all carbon derived fuels upon combustion. Hence, NG, SNG and biogas appear suitable mid-term fuels for mobility and electricity production in view of the replacement of fossil fuels. A particularly important issue related to $CO_2$ hydrogenation (the Sabatier reaction, $CO_2 + 4H_2 \rightarrow CH_4 + 2H_2O$) using Ni-based catalysts is the deactivation of the Ni active component over time due to particle growth. In this perspective, the research work described here summarizes our achievements with perovskite-type oxides as alternative support for Ni because of their self-regeneration property (SRP).

SRP was originally defined as the ability of a perovskite-type oxide lattice to segregate under reducing conditions a catalytically active element, i.e. typically a transition element that occupies the B-site (**1.**, Figure 5). Under oxidizing conditions the metal particles dissolve again in the lattice to restore the nanostructure in a reversible manner (**3.**). The process can in principle be repeated over several segregation-oxidation cycles.[26] This property was demonstrated primarily for Platinum group metals (PGM; Pt, Pd, Rh) in automotive applications where SRP was targeted to the stabilization of PGM nano-particles against sintering that causes catalyst deactivation over mileage.[27] Its demonstration for other transition metals involved in energy related catalytic processes is rare. Burnat et al. first assessed the SRP of La-Sr-Ti-Ni-O mixed oxides used as anode of solid oxide fuel cells (SOFC).[28] Figure 5 also shows an additional cycle beside SRP. If the oxidation temperature is not sufficiently high to revert all B-site atoms to the original oxidation and coordination states (**4.**), the material enters a cycle in which a supported catalyst is obtained. Upon reduction (**5.**), particles of the segregated active B-site metal are formed that have the catalytic properties of the particles generated in the SRP cycle. However, the key difference is that continuous redox cycling in the case of the supported catalyst cycle (**4.-5.**) can cause growth of the metal particles and deactivation over time. Here, we summarize the main results obtained in the effort to determine the SRP boundaries for $LaFe_{0.8}Ni_{0.2}O_3$. It is also demonstrated how the catalyst can be completely regenerated from deactivation by coke in a $CO_2$ methanation feed. The oxidation segment of the SRP cycle removes the carbon deposits and simultaneously regenerates the initial perovskite-type oxide structure in which Nickel occupies the B-site. In the following, this species is termed $Ni^{n+}_{oct}$, indicating that Ni is in higher oxidation state than +2 and in the octahedral coordination of the B-site. Coking is illustrated in Figure 5 as the result of reaction (**2.**) and as an accompanying phenomenon to Ni particle growth.

$LaFe_{0.8}Ni_{0.2}O_3$ was selected because of the ease of $LaFeO_3$ to deal with SRP,[26] while Ni is an active metal for various industrially relevant processes. This is not a typical combination for $CO_2$ hydrogenation.[29] In order to demonstrate the occurrence of SRP, it is necessary to verify i) that Ni can adopt the coordination environment of Fe in $LaFe_{0.8}Ni_{0.2}O_3$ ($Ni^{n+}_{oct}$), ii) that under selected conditions Ni can reversibly segregate and enter the $LaFeO_3$ lattice, and finally, iii) that this function remains at work during catalytic operation and ensures regeneration of the catalyst, e.g. upon coking. Hence, the catalyst would follow the self-regeneration cycle rather than the supported catalyst cycle of Figure 5.

In order to broaden the perspective of the potential SRP of Ni, the coordination environment of Ni was initially investigated in three perovskite-type oxides of different crystal structure (cubic, rhombohedral and orthorhombic) and composition ($La_{0.3}Sr_{0.55}TiO_3$, $LaCoO_3$ and $LaFeO_3$).[30] We have found previously in the case of Pd that incorporation of the active element in the perovskite-type structure depends on the composition of the mixed oxide, on the nature of the B-site elements and on the synthesis method.[31] Ni is capable of substituting Ti, Co and Fe without affecting the perovskite-type crystal phase. The Ni K-edge X-ray absorption near edge structure (XANES) spectra of the three materials exhibited different aspect and were also significantly dissimilar from that of NiO, including a shift to higher energy suggesting higher Ni oxidation state than +2. The XANES spectrum of $LaFe_{0.8}Ni_{0.2}O_3$ is shown in Figure 6a. The spectra included also common features, i.e. a weak pre-edge (ca. 8.334 keV; 1s→3d), an intense whiteline and a strong absorption dip (ca. 8.38 keV). While the coordination of Ni in these perovskite-type oxides (and in NiO) consists of six oxygen atoms in octahedral arrangement, the distinct spectra of the samples where Ni is intentionally dissolved in the mixed oxide originate from differences in the crystal symmetry and composition of the compounds. These characteristics affect the electron scattering from higher coordination shells than Ni-O. Ni-O bond distances obtained from the corresponding extended X-ray absorption fine structure data (EXAFS) were clearly shorter (1.95-1.98 Å) than those of NiO impregnated on the mixed oxides (2.03-2.07 Å). Hence, the coordination environment of Ni changes according to the composition and the crystal phase, and the oxidation state of Ni is higher than +2. The XANES spectra can be used as fingerprints of Ni within the perovskite-type structure and substituting the B-site element ($Ni^{n+}_{oct}$, Figure 5).

The reducibility of such materials is crucial for their utilization as precursors of catalysts obtained from segregation of Ni upon reduction at elevated temperatures. Repeated ramps in temperature programmed reduction by $H_2$ (TPR) up to 600°C interposed to oxidation at increasing temperature are a powerful tool to demonstrate the temperatures at which $Ni^{n+}_{oct}$ segregates upon reduction and is regenerated upon oxidation.[32] The TPR profiles of materials alike $LaFe_{0.8}Ni_{0.2}O_3$ containing $Ni^{n+}_{oct}$ and of $NiO/LaFeO_3$ containing supported NiO ($Ni^{2+}$) are very different (Figure 6c,d). Following the scheme of Figure 5, TPR of a material oxidized at a lower temperature (**4.**) than that required to regenerate $Ni^{n+}_{oct}$ contains the features of the reduction of $Ni^{2+}$ (peak at ca. 350°C). Oxidation at increasing temperature favours the disappearance of this feature in the subsequent TPR ramp and the appearance of the feature characteristic of $Ni^{n+}_{oct}$ (ca. 250°C). Despite the occupation of the B-site position in the selected mixed oxides, TPR revealed that the temperature of $Ni^{n+}_{oct}$ reduction is significantly different depending on the oxide composition and the crystal structure. It is also very different compared to that of NiO supported on the corresponding perovskite-type oxides, however without being consistently higher or lower. The order of reducibility decreased with increasing tolerance factor of the perovskite-type oxide (according to Ubic et al.[33]) in the order $LaFe_{0.95}Ni_{0.05}O_{3\pm\delta}$ (t= 0.942) > $LaCo_{0.95}Ni_{0.05}O_{3\pm\delta}$ (0.97) > $La_{0.3}Sr_{0.55}Ti_{0.95}Ni_{0.05}O_{3\pm\delta}$ (0.993). The large difference in temperature of Ni reduction in these compounds justifies their use in catalytic applications with vastly different operation temperatures. For example, $La_{0.3}Sr_{0.55}Ti_{0.95}Ni_{0.05}O_{3\pm\delta}$ is a self-regenerable SOFC anode material operating above 800°C.[34] $LaCo_{0.8}Cu_{0.2}O_3$ with similar low temperature reduction behaviour to $LaCo_{0.95}Ni_{0.05}O_{3\pm\delta}$ was used for preferential CO oxidation (< 400°C).[35]

Based on the alternated TPR-oxidation experiments, we identified 600°C as the lowest oxidation temperature required to obtain the same TPR profile of the initial material containing $Ni^{n+}_{oct}$, thus the temperature needed to fully regenerate $LaFe_{0.8}Ni_{0.2}O_{3\pm\delta}$.[36] Acceptable kinetics of oxidation become realistic from an experimental viewpoint at 650°C for 2 h. Importantly, under these conditions only Ni was found to segregate reversibly while no Fe reduction occurred (Figure 6b).[36] It should be noted that the conditions for reversible segregation-regeneration of $Ni^{n+}_{oct}$ are not valid above 20 mol% Ni because reduction starts to entail severe reduction of the host lattice to Fe metal and $La_2O_3$ that cannot be restored efficiently upon oxidation if not above 650°C and for longer time.

XANES spectra are also used to quantify the fraction of reduced Ni using linear combination fit of selected reference compounds. This is made easier by the above observation that the spectrum of $Ni^{n+}_{oct}$ is very different from that of NiO or supported NiO corresponding to $Ni^{2+}$. Hence, the spectra of $Ni^{n+}_{oct}$, once it is established that the sample contains exclusively or predominantly such species, the spectrum of NiO or better, of the corresponding supported NiO material, and the spectrum of Ni metal can be used. Segregated Ni formed nanoparticles of 10-20 nm of average size that are identified by scanning electron microscopy (SEM). Reduction at 600°C produces 35% of Ni metal in $LaFe_{0.95}Ni_{0.05}O_{3\pm\delta}$ and 50% in $LaFe_{0.8}Ni_{0.2}O_{3\pm\delta}$,[36] oxidation at 650°C restoring the initial XANES spectrum of $Ni^{n+}_{oct}$ and thus the original structure (Figure 6e).

The segregation-regeneration of $Ni^{n+}_{oct}$ can also be followed by X-ray diffraction (XRD). When Ni adopts the coordination environment of the B-site, the diffraction pattern shifts to either lower or higher angles depending on the composition of the perovskite-type oxide.[30] In the case of $LaFeO_3$, gradual contraction of the unit cell occurs with increasing Ni content compared to $LaFeO_3$ as a result of rotation of the $BO_6$ octahedra upon insertion of the small $Ni^{n+}_{oct}$ ion at $Fe^{3+}$ positions.[36] XRD is relatively silent to formation of Ni particles upon segregation as a consequence of their low density and small crystallite size. However, Ni segregation can be followed in terms of measurable expansion of the lattice that by comparison, is not visible in the patterns of the corresponding supported NiO materials because of the absence of $Ni^{n+}_{oct}$.[36]

At this point, consecutive oxidation-reduction cycles at the respective temperatures can be used to demonstrate that the size of the Ni particles after each reduction segment remain stable in terms of particle density and average size (Figure 6e). Importantly, this is not true for example for the $NiO/LaFeO_3$ analogue of $LaFe_{0.8}Ni_{0.2}O_{3\pm\delta}$ exposed to the same treatment. In this impregnated $LaFeO_3$ material, larger Ni particles than the initial ones were observed demonstrating the stabilizing effect of SRP on particle growth also for these materials that was at the origin of their work on SRP.[26]

$LaFeO_3$ produces very selectively CO from hydrogenation of $CO_2$ because of the lack of sites able to dissociate $H_2$. $CH_4$ production develops when Ni metal is present on $LaFeO_3$, either in Ni segregated from $LaFe_{0.8}Ni_{0.2}O_{3\pm\delta}$ or in reduced $Ni/LaFeO_3$. Stable catalysts are obtained after exposure to at least five consecutive reduction-oxidation cycles at the temperatures of SRP for this material. The larger particles of $NiO/LaFeO_3$ after the same treatment exhibit decreased catalytic activity. However, catalytic activity towards $CH_4$ should be further developed, a critical point being the amount of Ni that can be segregated upon reduction.

Finally, deactivation and regeneration of the catalyst need to be optimized as well in order to observe beneficial effects of SRP at lab sale. Ethylene ($C_2H_4$) was used to force deactivation of reduced $LaFe_{0.8}Ni_{0.2}O_3$ in isothermal measurements. After reaction in the presence of $C_2H_4$ in the $CO_2/H_2$ feed at 430°C, cutting off $C_2H_4$ resulted in the severe loss of $CH_4$ yield (Figure 6f). At this point, the catalyst surface was covered by coke filaments of graphitic nature (observed by SEM). According to the established SRP procedure, oxidation at 650°C removed completely the carbon deposits and the following reduction at 600°C produced the segregated Ni particles. After oxidation, the fluorescence XANES spectrum was identical to that of the unused sample, thus of $Ni^{n+}_{oct}$ suggesting that Ni reverted into the octahedral coordination simultaneous to the removal of carbon. Thus, SRP is at work also after extensive poisoning by carbon and is responsible to regenerate the poisoned catalyst.[37] Importantly, the size of the Ni particles and as a consequence the $CH_4$ yield remained stable along several consecutive reduction-coking-oxidation-reduction cycles suggesting the effective protection of Ni from particle sintering exerted by the oxidation segment of the SRP procedure irrespective of carbon coking. We recently showed that a similar regeneration can be obtained for SOFC anodes of the La-Sr-Ti-Ni-O type after poisoning with $H_2S$ at elevated temperatures,[34] thus demonstrating that this approach is of more general validity.

In conclusion, the boundaries for the SRP of $LaFe_{0.8}Ni_{0.2}O_3$ were established, which resulted in a stable Ni dispersion over various repeated reaction-coke poisoning-SRP cycles. SRP effectively suppresses growth of active metal particles but is a treatment not a preventive measure. Future work should be devoted to develop strategies i) to increase the $CH_4$ yield, for example upon substitution of the A-site,[38] and exploration of reaction conditions, ii) to improve the Ni reducibility and the efficiency of Ni usage, and finally, iii) to study the kinetics of reduction and oxidation including the use of spectroscopic methods to characterize the related structural variations.[39]

**Acknowledgements**

D.F. would like to thank Dr. P. Steiger for the passionate doctoral work and Dr. A. Heel (Zurich University of Applied Sciences, ZHAW) for the continuous fruitful cooperation, the Swiss National Science Foundation (SNF, project nr. 159568) and the Competence Center for Energy and Mobility (CCEM) for financial support, and beamline SuperXAS of the Swiss Light Source (SLS) at PSI for the support. E.F. gratefully thanks the SNF through its Ambizione Program, the National Competence Center in Research (NCCR) MARVEL, Innosuisse, the Swiss Competence Center for Energy Research (SCCER) Heat & Electricity Storage and PSI for financial contributions. D.P. gratefully thanks Dr. F. Haydous, Dr. W. Si, and Dr. M. Pichler for their contributions, the SNF and the NCCR MARVEL for financial support.

**Figure**

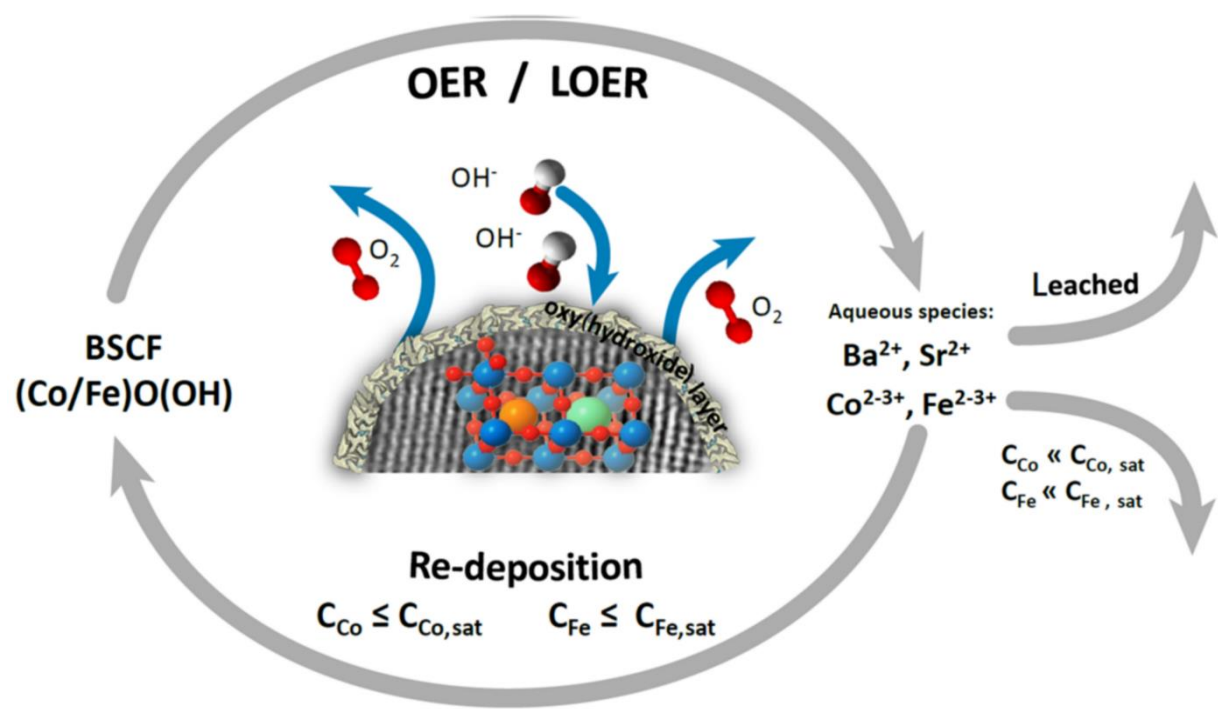

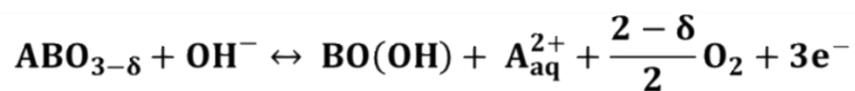

**Fig. 1.** OER/LOER and dissolution/redeposition mechanism leading to the formation of a self-assembled active surface layer on the surface of BSCF perovskite. Reprinted with permission from Chemistry of ACS Catalysis 8, 9765-9774 (2018). Copyright (2018) American Chemical Society.

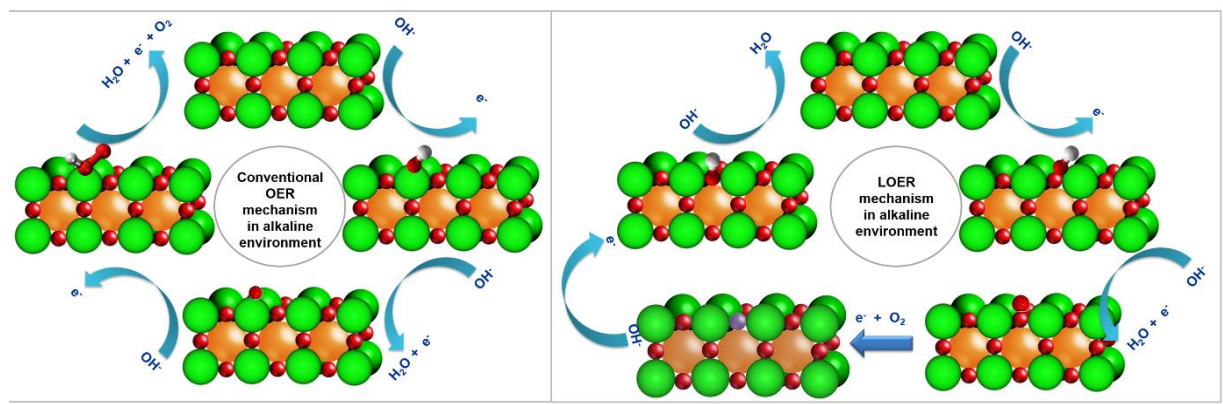

**Fig. 2.** On the left, the conventional OER mechanism involving proton electron transfers on the surface metal centers. On the right, the LOER mechanism proposed by Fabbri and Schmidt[4a] in alkaline environment where lattice oxygens represents the reactions sites, leading to lattice oxygen evolution and consequent formation of oxygen vacancies in the metal oxide lattice, which can be replenished by a final step by reacting with OH⁻ in the electrolyte. Four electrons are overall exchanged in the LOER but decoupled proton electron steps take place. A perovskite structure $ABO_3$ has been used as representative catalyst where the orange, green, and red spheres represent the A-site cations, B-site cations, and the oxygens, respectively. Reprinted with permission from Chemistry of ACS Catalysis 8, 9765-9774 (2018). Copyright (2018) American Chemical Society.

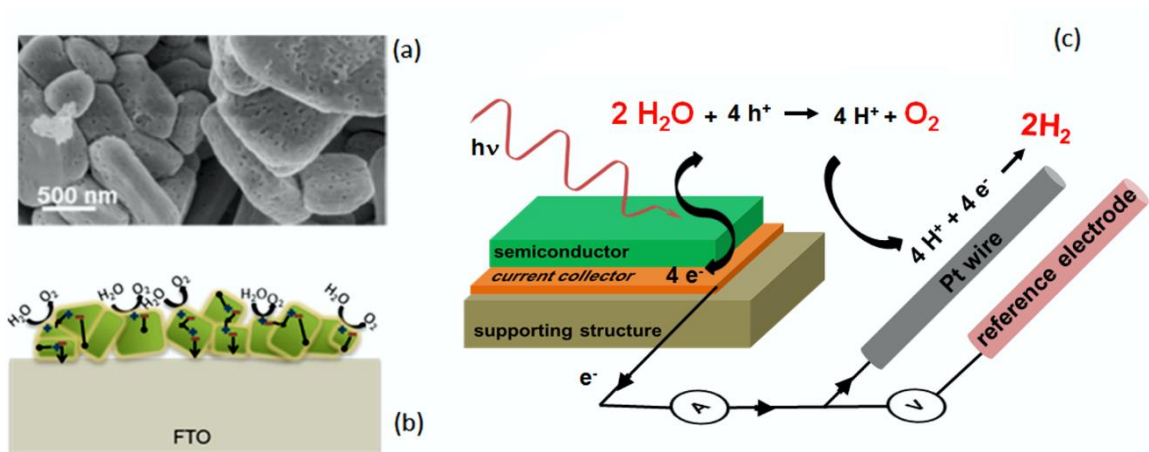

**Fig. 3.** (a) SEM micrograph of LaTiO$_2$N powder.[18a] (b) Schematic illustration of decorated powder deposited onto FTO glass. (Adapted with permission from ref. [18a]. Copyright (2019) American Chemical Society). (c) Sketch of the 3-electrode configuration for PEC characterization.

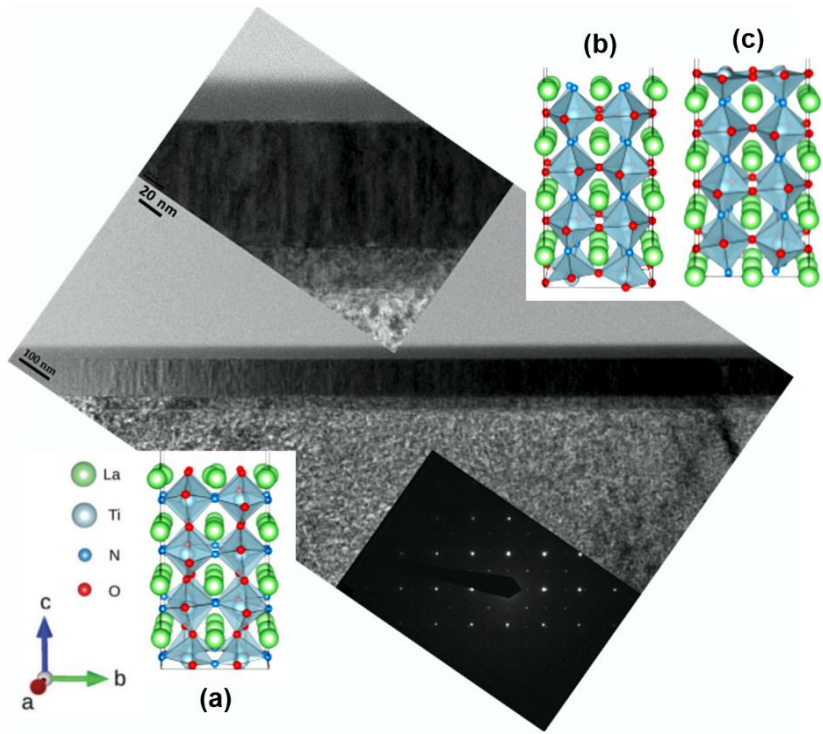

**Fig. 4.** Cross sectional TEM micrograph and SAED pattern of a (001)-oriented epitaxial LaTiO$_x$N$_y$ film grown on TiN-buffered (100)-oriented MgO substrate (courtesy of Dr. E. A. Müller, Electron Microscopy Facility, Paul Scherrer Institute). (a), (b) and (c) show respectively the LaO, La-N, and TiO$_2$ terminations of the (001) surface for the energetically preferential anion order. (Adapted with permission from ref. [21b]. Copyright (2017) WILEY-VCH Verlag GmbH & Co. KGaA, Weinheim).

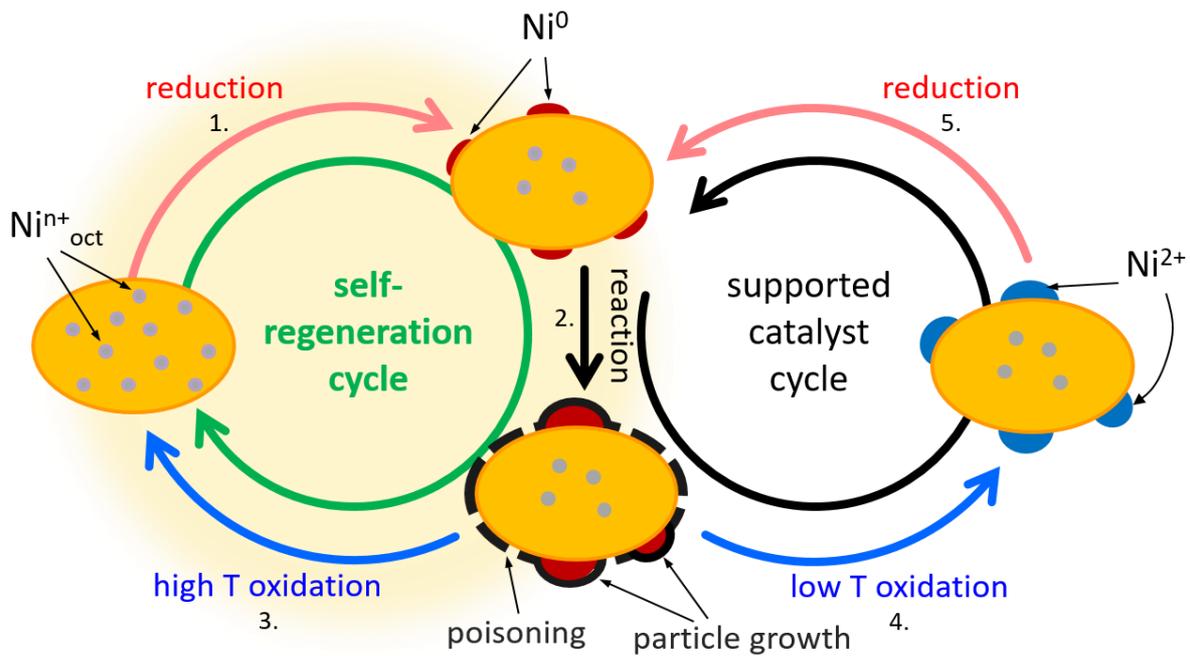

**Fig. 5.** The principle of self-regeneration of LaFe$_{0.8}$Ni$_{0.2}$O$_3$. Self-regeneration cycle: after reduction of the starting material (**1.**) and reaction (**2.**), calcination at sufficiently high temperature restores Ni in the perovskite-type oxide (**3.**). Supported catalyst cycle: calcination at lower temperature (**4.**) generates metal Ni particles supported on the Ni-deficient mixed oxide but could include a fraction of Ni within the perovskite-type oxide. After reduction (**5.**) to Ni metal, the supported catalyst (**2.**) can enter one of the two cycles.

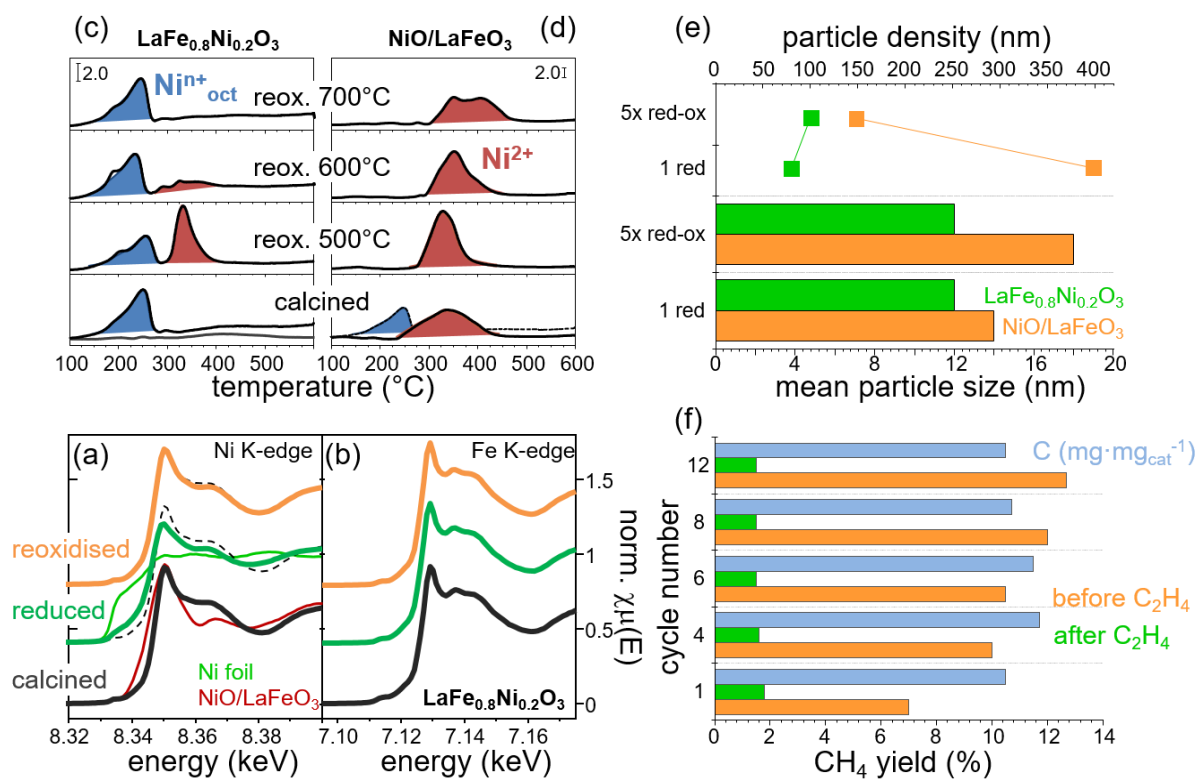

**Fig. 6.** Boundaries of the self-regeneration property (SRP). Structure of Ni in LaFe$_{0.8}$Ni$_{0.2}$O$_3$: (a) Ni K-edge and (b) Fe K-edge XANES spectra of LaFe$_{0.8}$Ni$_{0.2}$O$_3$ after various treatments as indicated. The spectra of Ni foil (Ni$^0$) and NiO/LaFeO$_3$ are also shown. The spectrum of Ni$^{n+}_{oct}$ in LaFe$_{0.8}$Ni$_{0.2}$O$_3$ is very different from that of octahedral Ni$^{2+}$ in NiO/LaFeO$_3$. The spectrum of calcined LaFe$_{0.8}$Ni$_{0.2}$O$_3$ is repeated in (a) for comparison purposes (dashed line). (c) TPR of LaFe$_{0.8}$Ni$_{0.2}$O$_3$ containing Ni$^{n+}_{oct}$ and (d) NiO/LaFeO$_3$ containing Ni$^{2+}$ after calcination and after oxidation at various temperatures. TPR of LaFeO$_3$ is shown for comparison in the bottom panel of (c). TPR of LaFe$_{0.8}$Ni$_{0.2}$O$_3$ is repeated in the bottom panel of (d) for comparison purposes. (e) SRP in terms of particle size and particle density: Ni particle size and density of LaFe$_{0.8}$Ni$_{0.2}$O$_3$ and NiO/LaFeO$_3$ after one reduction (600°C) and after five consecutive oxidation (650°C)-reduction (600°C) cycles. (f) Catalytic activity towards CO$_2$ methanation and stability towards repeated coke formation cycles by SRP: CH$_4$ yield before addition of C$_2$H$_4$ and after removing C$_2$H$_4$ in repeated methanation-coking-regeneration cycles on LaFe$_{0.8}$Ni$_{0.2}$O$_3$ in presence of 3000 ppm C$_2$H$_4$ (8 h) and corresponding amount of deposited carbon.